\def\ra{\rangle}
\def\la{\langle}
\def\ver{\arrowvert}

\documentstyle[pra,aps,preprint]{revtex}

\begin{document}
\title{Quantum mechanical Universal constructor}

\author{Arun K.\ Pati$^{(\dagger)}$ and Samuel L.\ Braunstein}
\address{
Informatics, Bangor University, Bangor LL57 1UT, UK\\
$^{(\dagger)}$Institute of Physics, Sainik School Post,\\
Bhubaneswar-751005, Orissa, India\\}

\date{\today}

\maketitle

\begin{abstract}
Arbitrary quantum states cannot be copied. In fact, to make a copy we must
provide complete information about the system. However, can a quantum
system self-replicate? This is not answered by the no-cloning theorem.
In the classical context, Von Neumann showed that a `universal
constructor' can exist which can self-replicate 
an arbitrary system, provided that it had access to instructions for making 
copy of the system. We question the existence of a universal constructor 
that may allow for the self-replication of an arbitrary 
quantum system. We prove that there is no deterministic universal quantum 
constructor which can operate with finite resources. Further, we delineate 
conditions under which such a universal constructor can be designed to operate 
dterministically and probabilistically.

\end{abstract}

%\begin{multicols}{2}

\vskip 1cm

The basis of classical computation is the Church-Turing thesis \cite{ch,at} 
which says that every recursive function can be computed algorithmically 
provided the algorithm can be executed by a physical process. 
However, fundamental physical processes are not governed by classical
mechanics, rather by quantum mechanical laws. The possibility of performing
reversible computation \cite{chb} and the fact that classical computers
cannot efficiently simulate quantum systems \cite{rf,pb} gave birth to the
concept of the quantum Turing machine \cite{david}.
This led to a flurry of discoveries in quantum computation \cite{rf1},
quantum algorithms \cite{bv,deu,ps,lkg}, quantum simulators 
\cite{seth}, quantum automaton \cite{albert} and programmable gate 
array \cite{nielsen}. 
%such as quantum complexity theory \cite{bv}, quantum algorithms for 
%testing global properties of a function \cite{deu}, factoring large 
%integers \cite{ps}, searching virtual data bases \cite{lkg} and general 
%quantum simulators \cite{seth}. 
%The question of designing a programmable quantum gate array 
%that can execute unitary operations on an arbitrary state with a fixed 
%circuit has been addressed \cite{nielsen}. 
In another development, von Neumann
\cite{vn} thought of an extension of the logical concept of a universal
computing machine which might mimic a living system. One of the hall-mark 
properties of a living system is its capability of self-reproduction.
He asked the question: Is it possible to design a machine that could be 
programmed to produce a copy of itself, in the same spirit that a Turing 
machine can be programmed to compute any function allowed by physical 
law. More precisely, he defined a {\em `universal constructor'} as a 
machine which can reproduce itself
if it is provided with a program containing its own description. The process
of self-reproduction requires two steps: first, the constructor has to
produce a copy of itself and second, it has to produce the program of how to
copy itself. The second step is important in order that the self-reproduction
continues, otherwise, the child copy cannot self-reproduce. When the
constructor produces a copy of the program, then it attaches it to the 
child copy and the process repeats. Unexpectedly, working with classical 
cellular automaton it was found that there is indeed a universal constructor 
capable of self-reproducing.

In a sense, von Neumann's universal constructor is a ``Turing test of life'' 
\cite{ca} if we attribute the above unique property to a living system, 
though there are other complex properties such as the ability to self-repair, 
grow and evolve. 
%Understanding of living systems and their properties 
%are fundamental objectives in all branches of science. 
From this perspective, the universal constructor is very useful model to 
explore and understand under what conditions a system is capable of 
self-reproducing (either artificially or in reality). 
%Here, we look for the existence of self-reproductive machines in 
%the quantum world. It may be emphasized that 
If one attempts to understand
elementary living systems as quantum mechanical systems in an information
theoretic sense, then one must first try to find out whether a {\em universal 
quantum constructor} exists. In a simple and decisive manner, we find that 
an all-purpose quantum mechanical constructor operating in a closed 
universe with a finite resource cannot exist.

The quantum world is fundamentally different in many respects than any 
classical world. There are many kinds of machines which are possible 
classically but impossible quantum mechanically. 
Wigner was probably the first
to address the question of replicating machines in the quantum world and found
that it is infinitely unlikely that such machines can exist \cite{wigner}.
It is now well known that the information content of 
a quantum state has two remarkable properties: first, it cannot be copied 
exactly \cite{wz,dd} and second, given several copies of an unknown state we
cannot delete a copy \cite{akp}. 
For example, if one could clone an arbitrary state then one could 
violate the Heisenberg uncertainty relation, and moreover, using non-local 
resources one could send signals faster than the speed of light \cite{dd}. 
In addition, non-orthogonal quantum states cannot be perfectly copied
whereas orthogonal quantum states can be. 
\cite{hy}. The extra information 
needed to make a copy must be as large as possible --- a 
recent result known as the stronger no-cloning theorem \cite{jozsa}. 
The no-cloning and the no-deleting principles taken together reveal 
some kind of `permanence' of quantum information.

First, we observe that merely copying of information is {\em 
not the self-replication}. Therefore, with the quantum mechanical toolkit, 
we must formalize the question of a self-replicating machine.
Let $A$ be a universal construction machine. If $\ver \Psi \ra$ is
the state of a species that would self-replicate, then by furnishing
a suitable description of the instructions $U$ to produce $\ver\Psi\ra$
(in accordance with the stronger no-cloning
theorem we have to supply the full information about $\ver \Psi \ra$) then
$A$ will construct a copy of $\ver \Psi\ra$.
However, this $A$ is not yet self-reproducing, because $A$ has produced a copy
of $\ver \Psi \ra$ but without a copy of $U$. If we add the description of
$U$ to itself it will not solve the problem, as this would lead to infinite
regression. In von Neumann's spirit we can imagine that
there exists an additional quantum system $B$ that stores the instructions 
$U$ and can make a copy of them.
Another ancillary system $C$, called the control unit, will insert 
the copy of $U$ into $\ver \Psi \ra$ and then will separate from 
the composite system $A+B$. 

A quantum mechanical universal constructor may be completely specified
by a quadruple ${\cal U}{\cal C}= (\ver \Psi\ra, \ver P_U \ra, \ver C
\ra, \ver \Sigma \ra)$, where $\ver \Psi\ra \in {\cal H}^N $ is the state of 
the  (artificial   or  real)  living  system   that  contains  quantum
information to  be self-replicated, $\ver  P_U \ra \in {\cal  H}^K$ is
the  program state  that contains  instructions to  copy  the original
information, i.e., the  unitary operator $U$ needed to  copy the state
$\ver \Psi \ra$ via $U(\ver \Psi \ra \ver 0\ra)= \ver \Psi \ra\ver \Psi 
\ra$ is encoded in the program state, $\ver C \ra$ is the state of the 
control unit, and $\ver \Sigma \ra=\ver 0 \ra
\ver 0 \ra \cdots \ver 0 \ra \in {\cal H}^M$ is a collection of blank 
states onto which the information and the program will be copied.  Let
there be  $n$ number of blank  states and if  each of $\ver 0  \ra \in
{\cal H}^N$,  then the dimension of  the blank state  Hilbert space is
$M=  N^n$. Without loss  of generality  we may  assume that  the blank
state $\ver 0 \ra$ may belong to a Hilbert space of dimension equal to
$N$. It  is assumed  that a  {\em finite} string  of blank  states are
available  in the environment  in which  the universal  constructor is
operating (they are analogous  to the low-entropy nutrient states that
are required  by a real  living system). The justification  for finite
number of  such states comes  from the fact  that in the  universe the
total  energy and  negative entropy  available at  any time  is always
finite \cite{wigner}. To  copy the program state the  machine uses $m$
blank states  in one generation, so  $K= N^m$. Thus $M$  is finite but
$M\gg N,K$.  The initial state  of the universal constructor  is $\ver
\Psi\ra  \ver  P_U \ra  \ver  C \ra  \ver  \Sigma  \ra$.  A  universal
constructor will be  said to exist if it can  implement copying of the
original and  the stored  program by a  fixed linear  unitary operator
${\cal L}$ acting on the combined Hilbert space of the input, program,
control   and  $(m+1)$   blank  states   that  allows   the  following
transformation
\begin{equation}
{\cal L}(\ver \Psi\ra \ver 0 \ra \ver P_U \ra \ver 0 \ra^{\otimes m} 
\ver C \ra) \ver 0\ra^{\otimes n-(m+1)}=
\ver \Psi\ra \ver P_U \ra {\cal L}(\ver \Psi\ra  \ver 0\ra \ver P_U \ra
\ver 0\ra^{\otimes m} \ver C' \ra) \ver 0\ra^{\otimes n-2(m+1)},
\end{equation}
where $\ver C'\ra$ is the final state of the control unit.
It is worth emphasizing that (1) is not a cloning transformation. It is a 
{\em recursively defined transformation} where the fixed unitary operator 
${\cal L}$ acts on the initial (parent) configuration and the same acts 
on the final (child) configuration after the copies have been produced. 
This definition is required in order that the self-replication proceeds in an 
{\em autonomous} way until the blank states are exhausted. The fixed unitary 
operator will not act on the child configuration unless $(m+1)$ nutrient 
states are available in the universe. Once the transformation is complete, the 
control unit separates the original information from the program states 
(parent information) so that the off-spring exists independently.
(i.e. there is no quantum entanglement between the parent and the child 
information). It then continues to self-reproduce.

If such a universal constructor exists, then when it is fed with another
state $\ver \Phi \ra$ and a suitable program $\ver P_V \ra $ to create it via 
$V(\ver \Phi \ra \ver 0\ra)= \ver \Phi \ra\ver \Phi \ra$ 
then it will allow the transformation
\begin{equation}
{\cal L}(\ver \Phi\ra \ver 0 \ra \ver P_V \ra \ver 0 \ra^{\otimes m}\ver C \ra)
\ver 0\ra^{\otimes n-(m+1)}=
\ver \Phi\ra \ver P_V \ra {\cal L}(\ver \Phi\ra  \ver 0\ra \ver P_V \ra
\ver 0\ra^{\otimes m} \ver C'' \ra) \ver 0\ra^{\otimes n-2(m+1)}.
\end{equation}
If such a machine can make a copy of any state along with its program 
in a unitary manner, then it must preserve the inner product. This 
implies that we must have
\begin{equation}
\la \Psi \ver \Phi \ra \la P_U \ver P_V \ra=
\la \Psi \ver \Phi \ra^2 \la P_U \ver P_V \ra^2 \la C' \ver C'' \ra
\end{equation}
holds true. However, 
the above equation tells us that the universal constructor can exist only 
under two conditions, namely, (i) either $\la \Psi \ver \Phi \ra =0$ 
and $\la P_U \ver P_V \ra \not= 0$ or (ii) $\la \Psi \ver \Phi \ra \not=0$ 
and $\la P_U \ver P_V \ra=0$. The first condition suggests that for 
orthogonal states as the carrier of information, there is no restriction 
on the program state. This means that with a finite dimensional program 
state and finite number of blank states orthogonal states can self-replicate. 
Such a universal constructor can exist with a finite resources. This 
corresponds to the realization of a classical universal constructor, 
and is consistent with von Neumann's thesis, that a self-reproducing general 
purpose machine can exist, in principle, in a deterministic universe 
\cite{vn}. However, the second condition tells us that for non-orthogonal 
states, the program states have to be
orthogonal. This means that to perfectly self-replicate a collection of
non-orthogonal states $\{\ver \Psi_i\ra\}$ together with their program
states $\{\ver P_{U_i} \ra\}, (i=1,2, \ldots )$ one requires that the set
$\{\ver P_{U_i} \ra\}$'s should be orthogonal. Since an arbitrary state 
such as $\ver \Psi \ra = \sum_i \alpha_i \ver i \ra$ with the complex numbers 
$\alpha_i$'s varying continuously can be viewed as an infinite collection 
of non-orthogonal states (or equivalently the set of non-orthogonal states 
for a single quantum system is infinite, even for a simplest two-state system 
such as a qubit), one requires an infinite-dimensional program state to 
copy it.  In one generation of the self-replication the number of blank states
used to copy the program state is $m= \log_2 K/\log_2 N$ and when 
$K\rightarrow \infty$ the nutrient resource needed also becomes infinite. 
As a consequence, to copy an infinite-dimensional Hilbert space program 
state one needs an infinite collection of blank states to start with.
Furthermore, the number of generations 
$g$ for which the self-reproduction can occur with a finite nutrient resource 
is $g= \log_2 M/(\log_2 K  N)$. When $K$ becomes infinite, then there can be 
no generations supporting self-reproduction.
Therefore, we surmise that {\em with a finite-dimensional program state 
and a finite nutrient resource there is no deterministic universal constructor 
for arbitrary quantum states.}  
However, if one is interested 
in self-replication of a finite number $K$ of the non-orthogonal states 
with $K$ the dimension of the program Hilbert space, then it may be possible 
to design a universal constructor with finite resources. Because, one may 
find $K$ mutually orthogonal program states that span the program Hilbert 
space.

One may ask is it not possible to rule out the nonexistence of
deterministic universal constructor from no-cloning principle? The
answer is `no' for two reasons. First, a simple universal cloner is {\em not}
a universal constructor. Second, in a universal constructor we provide 
the complete specification about the input state, hence it should have been 
possible to self-reproduce, thus reaching an opposite conclusion! 
Even though Eq.(3) may look the same to what one gets in the cloning 
operation, the transformation defined in Eq.(1) is not. One similarity for 
example, is that if one stores the information in an orthonormal basis states 
$\{ \ver \Psi_i \ra \}$, $(i=1,2,...N$), then the cloning and the 
self-replication both are allowed. 
The surprising and remarkable result is that when we ask to self-replicate any 
arbitrary living species, then it cannot. 
By providing complete information of a quantum system, one may think that 
it should be able to self-replicate. Because 
when we {\em know} the state completely then we can design a program to 
copy the state.
%, i.e., when we allow the program (unitary operator) to depend on the 
%state being copied then we can copy any state. 
If a universal quantum constructor exist then, the program is supposed to 
contain the description of the species, 
i.e. the information about $\ver \Psi \ra$.
So it should have been able to self-replicate. But 
The perplexity of the problem lies when we allow the program to be 
copied. If it has to self-replicate then it violates unitarity of 
quantum theory. 

%It is interesting to note that though quantum no-cloning principle is valid
%the first requirement is not a problem. But the second requirement
%is not allowed due to linear nature of operators in quantum theory. 
%So even if there exist a constructor (containing a program of its own 
%description) which is capable of reproducing itself, the child
%constructor cannot, i.e., it will be sterile. 
%The impossibility of designing such a constructor unitarily arises from the
%problem of self-referencing, in that the constructor must make a copy
%of the program doing the construction.

This result may have immense bearing on explaining life based on quantum
theory. One may argue that after all if everything comes
to the molecular scale then there are variety of physical actions
and chemical reactions which might be explained by the basic laws
of quantum mechanics. However, 
%our result shows that, perhaps, one does 
%not need quantum mechanics to explain the self-reproducing nature of the 
%living system. Because, the self-reproducible information must be `classical' 
%the replication of DNA in a living cell, for example, can be understood 
%purely by classical means. 
if one applies quantum theory, then as we have proved quantum mechanical 
living organism cannot self-replicate. Interpreting differently, we might 
say that the present structure of quantum theory cannot model a living 
system as it fails to mimic a minimal living system. {\em For quantum 
mechanizing a living system seems to be an impossible task}. If that holds 
true, then this conclusion is going to have rather deep implication on 
our present search for ultimate laws of nature encompassing physical and 
biological world. On the other hand, because the self-reproducible information
must be `classical' the replication of DNA in a living cell can be 
understood purely by classical means. Having said this, our result does not 
preclude the possibility that quantum theory might play a role in explaining 
other features of the living systems
\cite{penrose}. For example, there is a recent proposal that quantum
mechanics may explain why the living organisms have four nucleotide bases and 
twenty amino acids \cite{patel}. It has been also reported that the game of 
life can emerge in the semi-quantum mechanical context \cite{fd}.

Though we have ruled out the existence of deterministic universal constructor
with finite resource, one can construct probabilistic universal constructor
for non-orthogonal species states $\{\ver \Psi_i \ra \}$ with certain 
probability of success, given a finite
dimensional program state $\{ \ver P_{U_i} \ra \}$ and a finite collection 
of blank states. 
%We can state the main result without the proof that the following 
%equation describes a probabilistic universal constructor. 
It is given by
 \begin{eqnarray}
&& {\cal L}(\ver \Psi_i \ra \ver 0 \ra \ver P_{U_i} \ra \ver 0 \ra^{\otimes m}
\ver C \ra \ver M \ra) \ver 0\ra^{\otimes n-(m+1)} = \nonumber\\
&& \sqrt{p_i} \ver \Psi_i \ra \ver P_{U_i} \ra {\cal L}(\ver \Psi_i\ra  
\ver 0\ra \ver P_{U_i} \ra
\ver 0\ra^m \ver C' \ra \ver M' \ra) \ver 0\ra^{ \otimes n-2(m+1)}
+  \sqrt{1-p_i} \ver X_i \ra,
\end{eqnarray}
where $p_i$ is the probability of success that the universal constructor works,
$\ver M\ra$ is the initial probe state and $\ver M' \ra$ is the final probe
state whose measurement can tell that it really succeeds,
and $\ver X_i \ra$ is the failure component of the whole constructor. It can
be proved that the above transformation can exists if and only if the set 
$\{\ver \Psi_i\ra \ver P_{U_i} \ra \}$ is linearly independent \cite{dg1}.
This implies that the quantum species states need not be necessarily linearly 
independent. It is only sufficient to have that condition satisfied. 
%The proof of this statement will be in the same line as that of the
%existence of probabilistic cloning machines \cite{dg1,akp}. 
The error in the probabilistic self-replication process of two
non-orthogonal states $\ver \Psi_i \ra$ and $\ver \Psi_j \ra$ is bounded by
\begin{equation}
f_{ij} \ge \frac{|\la \Psi_i\ver \Psi_j \ra| |\la P_{U_i}\ver P_{U_j} 
\ra| }
{1+|\la \Psi_i\ver \Psi_j \ra| |\la P_{U_i}\ver P_{U_j} \ra|}.
\end{equation}
From the above it is clear that there is no error introduced in the
self-replicating process of any two orthogonal states, even if the program 
states are not orthogonal.

Implications of our results are multifold for physical and biological 
sciences. It is beyond doubt that progress in the burgeoning area of 
quantum information technology can lead to revolutions in the machines that
one cannot think of at present. If a quantum mechanical universal constructor 
would have been possible, future technology would have allowed quantum 
computers to self-replicate themselves with little or no human input. 
That would have been a complete autonomous device --- a truly marvelous 
thing. However, a deterministic universal constructor with a finite 
resources is impossible in principle. One has to look for probabilistic 
universal constructors which can self-replicate with only limited 
probability of success, similar to probabilistic cloner \cite{dg1}. 
This could still have great implications for the future. 
With complete specification such a machine could construct copies 
based on its own quantum information processing devices. Future lines 
of exploration may lead to the design of approximate universal constructors 
in analogy with approximate universal quantum cloners \cite{bh}. 

How life emerges from inanimate quantum objects has been a conundrum 
\cite{sch,wme,gjc}. What we 
have shown here is that quantum mechanics fails to mimic a self-reproducing 
unit in an autonomous way. Nevertheless, if one allows for errors in 
self-replication, which actually do occur in real living systems, then 
an approximate universal constructor should exist. Such a machine would 
constitute a quantum mechanical mutation machine. It would be important to see
how variations in `life' emerge due to the errors in 
self-replication. From this perspective, if quantum mechanics is the final 
theory of nature, our result indicates that 
%negative in the sense that 
the information stored in a living organism 
are copied imperfectly and the error rate may be just right in order for 
mutation to occur to drive Darwinian evolution. In addition, one could study 
how the quantum evolution of species leads to an increase in the level of 
complexity in living systems. Since understanding these 
basic features of life from quantum mechanical principles is a fundamental 
task, we hope that the present result is a first step in that direction, 
and will be important in the areas of quantum information, artificial life,
cellular automaton, and last but not least in the
biophysical science.

\vskip 1cm
{\bf Acknowledgments:} 
We thank C. Fuchs for bringing the paper of E. P. Wigner to our attention. 
AKP wishes to thank C. H. Bennett, S. Lloyd, I. Chuang and D. Abbott 
for useful discussions. 
This work started during AKP's stay at Bangor University during 1999 
and was completed during a visit to MSRI, University of Berkeley, in 2002. 
SLB currently holds a Wolfson-Royal Society Research Merit Award.

%\end{references}

\noindent

%\end{multicols}


\begin{thebibliography}{99} 
%\begin{references}





\bibitem{ch} A. Church, 
%An unsolvable problem of elementary number theory. 
 Am. J. Math. {\bf 58} 345-363 (1936).

\bibitem{at} A. M. Turing, 
%On computable numbers with an application to the Entscheidungs problem. 
 Proc. Lond. Math. Soc. {\bf 42}, 230-265 (1936).

\bibitem{chb} C. H. Bennett, 
%Logical reversibility of computation. 
IBM Journal of Research and Development, {\bf 17}, 525-532 (1973).

\bibitem{rf} R. Feynman, 
%Simulating physics with computers.
 Int. J. Theo. Phys. {\bf 21}, 467-488 (1982).


\bibitem{pb} P. Benioff, 
%Quantum mechanical Hamiltonian models of Turing machines that 
%dissipates no energy.
 Phys. Rev. Lett. {\bf 48}, 1581-1585 (1982).

\bibitem{david}  D. Deutsch, 
%Quantum theory, the Church-Turing principle
%and the universal computers.
Proc R. Soc. Lond A {\bf 400}, 97-117 (1985).

\bibitem{rf1} R. Feynman, 
%Quantum mechanical computers. 
 Found. Phys. {\bf 16}, 507-531 (1986).

\bibitem{bv} E. Bernstein, and U. Vazirani, 
%Quantum complexity theory.
Proc. 25th Annual ACM Symposium on Theory of Computing, 11-20 (1993).

\bibitem{deu} D. Deutsch, \& R. Jozsa, 
%Rapid solutions of problems by quantum computation. 
 Proc. R. Soc. London A {\bf 449}, 553-558 (1992).

\bibitem{ps} P. Shor, 
%Polynomial-time algorithms for prime factorization
%and discrete logarithms on a quantum computer. 
 SIAM J. Comput. {\bf 265}, 1484-1509 (1997)

\bibitem{lkg} L. K. Grover, 
%Quantum mechanics helps in searching a needle in a haystack. 
 Phys. Rev. Lett. {\bf 79}, 325-328 (1997).

\bibitem{seth} S. Lloyd, 
%Universal quantum simulators. 
Science {\bf 263}, 1073-1078 (1996).

\bibitem{albert} D. Z. Albert, 
%On quantum mechanical automata. 
 Phys. Lett. A  {\bf 98}, 249-252 (1983).

\bibitem{nielsen} M. A. Nielsen, \& I. Chuang, 
%Programmable quantum gate arrays. 
Phys. Rev. Lett. {\bf 79}, 321-324  (1997). 


\bibitem{vn} J. von Neumann, The Theory of Self-Replicating Automata.
{\it University of Illinois Press, Urbana, IL} (1966) (work by
 J. von Neumann in 1952-53).


\bibitem{ca} C. Adami, 
%On modeling life. 
 Artificial Life {\bf 1}, 429-438 (1995).

\bibitem{wigner} E. P. Wigner, The Probability of the Existence of a 
Self-Reproducing unit. {\it The Logic of Personal Knowledge: Essays Presented
 to Michael Polany on his Seventieth Birthday} (Routledge \& Kegan Paul,
London) 231-238 (1961).

\bibitem{wz} W. K. Wootters, \& W. H. Zurek, 
%A single quantum cannot be cloned. 
Nature {\bf 299}, 802-803 (1982).

\bibitem{dd} D. Dieks, 
%Communication by EPR devices. 
Phys. Lett. A {\bf 92}, 271-272 (1982).

\bibitem{akp} A. K. Pati, \& S. L. Braunstein, 
%Impossibility of deleting an unknown quantum state. 
 Nature {\bf 404}, 164 (2000).

%\bibitem{ng} Gisin, N. Quantum cloning without signalling. {\it Phys. Lett. A}
%{\bf 242}, 1-3 (1998).

\bibitem{hy}  H. P. Yuen, 
%Amplification of quantum states and noiseless photon amplifiers. 
Phys. Lett. A {\bf 113}, 405-407 (1986).

\bibitem{jozsa} R. Jozsa, 
%A stronger no-cloning theorem. 
Quant. Ph., 0204153 (2002).

\bibitem{penrose} R. Penrose, Shadows of the Mind. {\it Oxford University
Press, Oxford}, (1994).

\bibitem{patel} A. Patel, 
%Why genetic information processing could have a quantum basis. 
J. of Bioscience {\bf 26}, 145-151 (2001).

\bibitem{fd} A. P. Flitney, \& D. Abbott, 
%A semi-quantum version of the game of life. 
 Quant. Ph., 0208149 (2002).

\bibitem{dg1}  L. M. Duan, \&  G. C. Guo, 
%Probabilistic cloning and identification of linearly independent states. 
Phys. Rev. Lett. {\bf 80}, 4999-5002 (1998). 

%\bibitem{arun} A. K. Pati, Quantum superposition of multiple clones and the
%novel cloning machine. {\it Phys. Rev. Lett.} {\bf 83}, 2849-2852 (1999).

\bibitem{bh} V. Bu$\check{z}$ek, \&  M. H. Hillery, 
%Quantum copying: Beyond the no-cloning theorem. 
Phys. Rev.A {\bf 54}, 1844-1852 (1996).

\bibitem{sch} E. Schr{\"o}dinger,  What is life . 
{\it Cambridge University Press, London}, (1944).
 
\bibitem{wme} W.\ M.\ Elsasser, The Physical Foundation of Biology,
{\it Pergamon Press, London}, (1958).

\bibitem{gjc} G. J. Chaitin, 
%To a mathematical definition of life.  
ACM SICACT News {\bf 4}, 12-18 (1970).

\end{thebibliography}
\end{document}